\documentclass[a4paper]{jpconf}
\usepackage{graphicx,psfrag,cite}
\begin{document}
\title{Engineered quantum tunnelling in extended periodic potentials}

\author{Sandro Wimberger\footnote[1]{Present address: CNISM and Dipartimento di Fisica del Politecnico, 
Corso Duca degli Abruzzi 24, I-10129 Torino}, Donatella Ciampini, Oliver Morsch, Riccardo Mannella, and 
Ennio Arimondo}
\address{CNR-INFM and Dipartimento di Fisica `Enrico Fermi', Largo Pontecorvo 3, I-56127 Pisa}

\ead{sandro.wimberger@polito.it}

\begin{abstract}
Quantum tunnelling from a tilted, but otherwise periodic potential is studied.
Our theoretical and experimental results show that, by controlling the system's parameters,
we can engineer the escape rate of a Bose-Einstein condensate to an exceptional degree.
Possible applications of this atom-optics realization of the open Wannier-Stark system are 
discussed.
\end{abstract}

\section{Introduction}
\label{intro}\medskip

One of the first manifestations of quantum mechanics was radioactive decay, in which -- according
to Gamov's theory -- a particle can overcome a potential barrier because of an even exponentially small
tail of its spatial wave function at the unbounded side of the barrier \cite{qm}. Besides this static
problem of over-the-barrier tunnelling, also dynamical tunnelling mechanisms are well-know today \cite{heller}. The
latter situation is found in systems where particles can escape from dynamical barriers (such as
regular elliptic islands in classical phase space) on the grounds of dynamically induced coupling 
processes (see, e.g., \cite{dyntun} for realistic examples).

Quantum tunnelling has found many technological applications, such as, for instance, in scanning 
tunnelling microscopes \cite{binnig} and in superconducting squid devices \cite{koelle}. Arguably
the mostly used application of tunnelling is present in tunnelling diodes and related integrated
semiconductor devices which go back to the pioneering work of Leo Esaki \cite{esaki1974}. The latter
also proposed to exploit resonantly enhanced tunnelling (RET) for technical use, and since the 1970's much
progress has been made in producing artificial super-lattice structures \cite{esaki1986}, in which
RET of fermionic quasiparticles could be demonstrated \cite{chang,chang_book}.

\begin{figure}[ht]
\includegraphics[width=18pc]{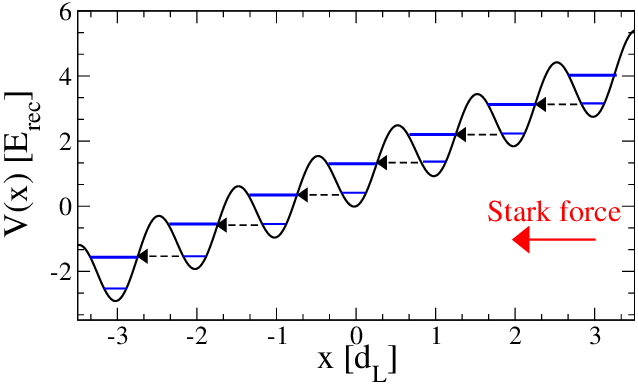}\hspace{2pc}
\begin{minipage}[b]{18pc}\caption{\label{fig:1} RET: tunnelling of atoms from a tilted lattice is
resonantly enhanced when the energy difference between the levels within an energetically equidistant
Wannier-Stark ladder (e.g. made up of the lower levels in all the wells) 
matches the distance between the energy levels in the wells.}
\end{minipage}
\end{figure}

Here we report the realization of RET using Bose-Einstein condensates which are held in optically
produced potentials (``optical lattices''). The counter-propagating beams creating the lattice can be
easily and in a controlled manner accelerated with respect to each other such as to mimic an additional 
static linear potential in the moving frame of reference \cite{peik}. Tunnelling in this 
Wannier-Stark system occurs between the quantised energy levels (the Wannier-Stark levels) in various
wells of the potential. An example of RET between next-nearest potential wells is illustrated in
Figure~\ref{fig:1}. In such a situation of tunnelling between energetically degenerate levels, the
escape rate can be varied by orders of magnitude by a slight change of the static tilting force 
\cite{GKK}.

The part of the condensate which has tunnelled through the barrier is in good approximation just
accelerated to infinity by the static force. As a consequence, we are dealing with an open, i.e.
non-hermitian decay problem \cite{GKK,WSM}, for which theoretical interest has recently revived within the
more general context of avoided level crossings in the complex energy plane \cite{GKK,avron,WGWK}.

In contrast to solid-state realizations of RET (see, e.g., 
\cite{chang_book}), our system with ultracold atoms offers a near-perfect
and easily reproducible control over parameters (the tilting force and the potential depth), the geometry
(tunnelling along an almost arbitrarily long array of potential wells and the condensate density), and
over the initial conditions (energy and quasimomentum of the prepared condensate). 
Decay rates for non-interacting cold atoms were measured for the Wannier-Stark system in a small interval
of the Stark force \cite{bha}, and a systematical study of RET with such cold atomic systems 
has been proposed in ref.~\cite{GKK}. Here we show that RET is accessible
to modern experiments with ultracold atomic gases, not only for the case of independently
moving atoms (described by the usual ``linear'' Schr\"odinger equation) but even in the presence of
a nonlinearity (with a mean-field nonlinear term mimicking the interatomic interactions).

\section{The nonlinear Wannier-Stark system and experimental results}
\label{results}\medskip

\begin{figure}[ht]
\includegraphics[width=12.5cm]{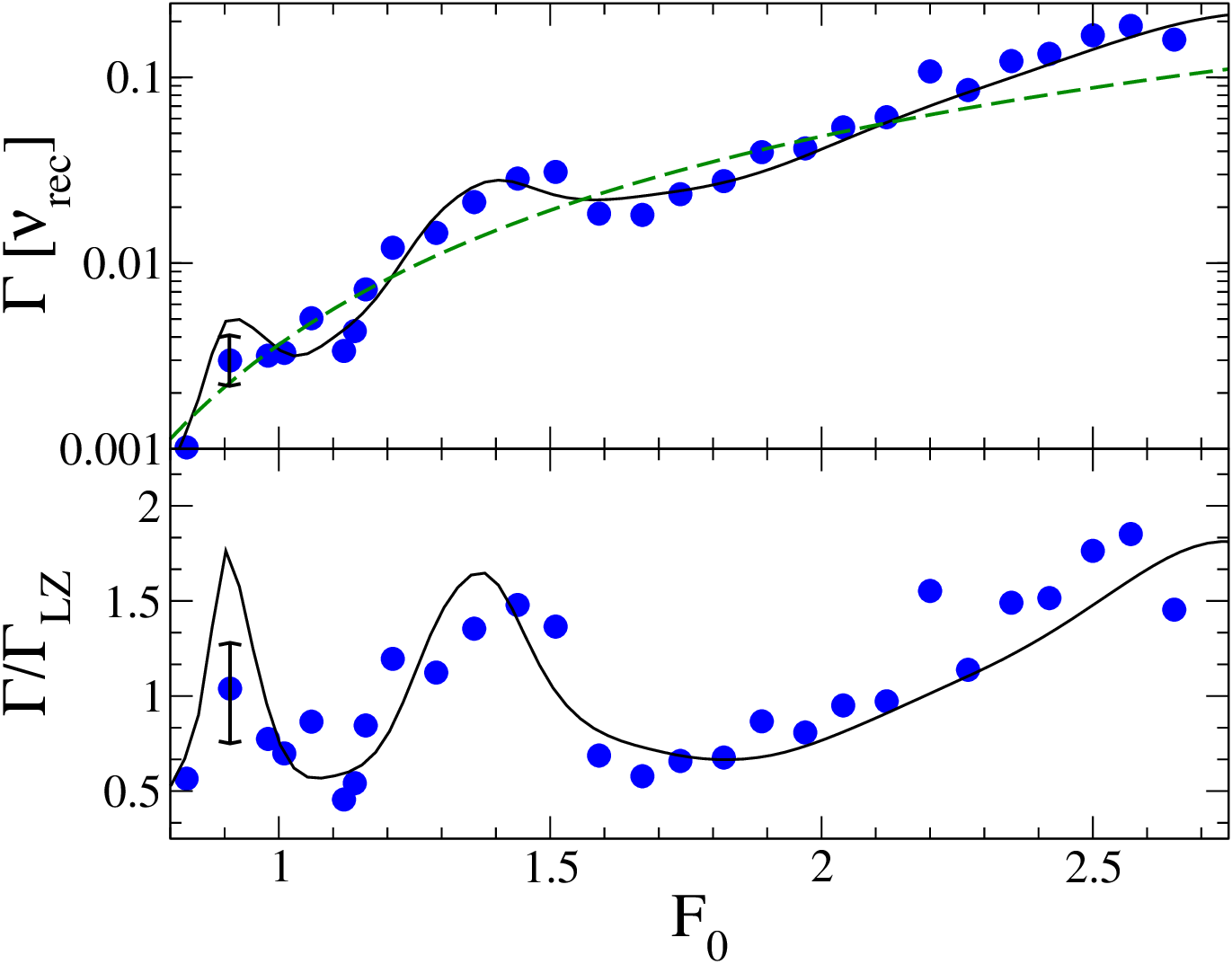}
\caption{\label{fig:2} 
Upper panel: measured Wannier-Stark tunnelling rates (\fullcircle) for $g \approx 0$ in a lattice
of depth $V_0 \approx 3.5$. The full line shows the theoretical expectation obtained by the methods
described in \cite{GKK,WSM}, while the dashed curve represents the  Landau-Zener prediction  
$\Gamma_{\rm LZ}$. Lower panel: the ratio of the theoretical (\full) and the measured (\fullcircle) 
rates and $\Gamma_{\rm LZ}$. Even if the modulation here is relatively small 
due to the chosen small lattice depth, we note the extention of the measured rates over two orders of 
magnitude in the upper panel. The nicely resolved central peak around $F_0\approx 1.35$ corresponds to
RET between second-nearest potential wells, while the left one at smaller $F_0\approx 0.5$ is a consequence
of RET between third-nearest well. The shoulder on the right shows signatures of the next-nearest neighbour
RET peak around $F_0 \approx 2.7$, corresponding to the situation sketched in figure~\ref{fig:1}.
}
\end{figure}

\begin{figure}[ht]
\includegraphics[width=12.5cm]{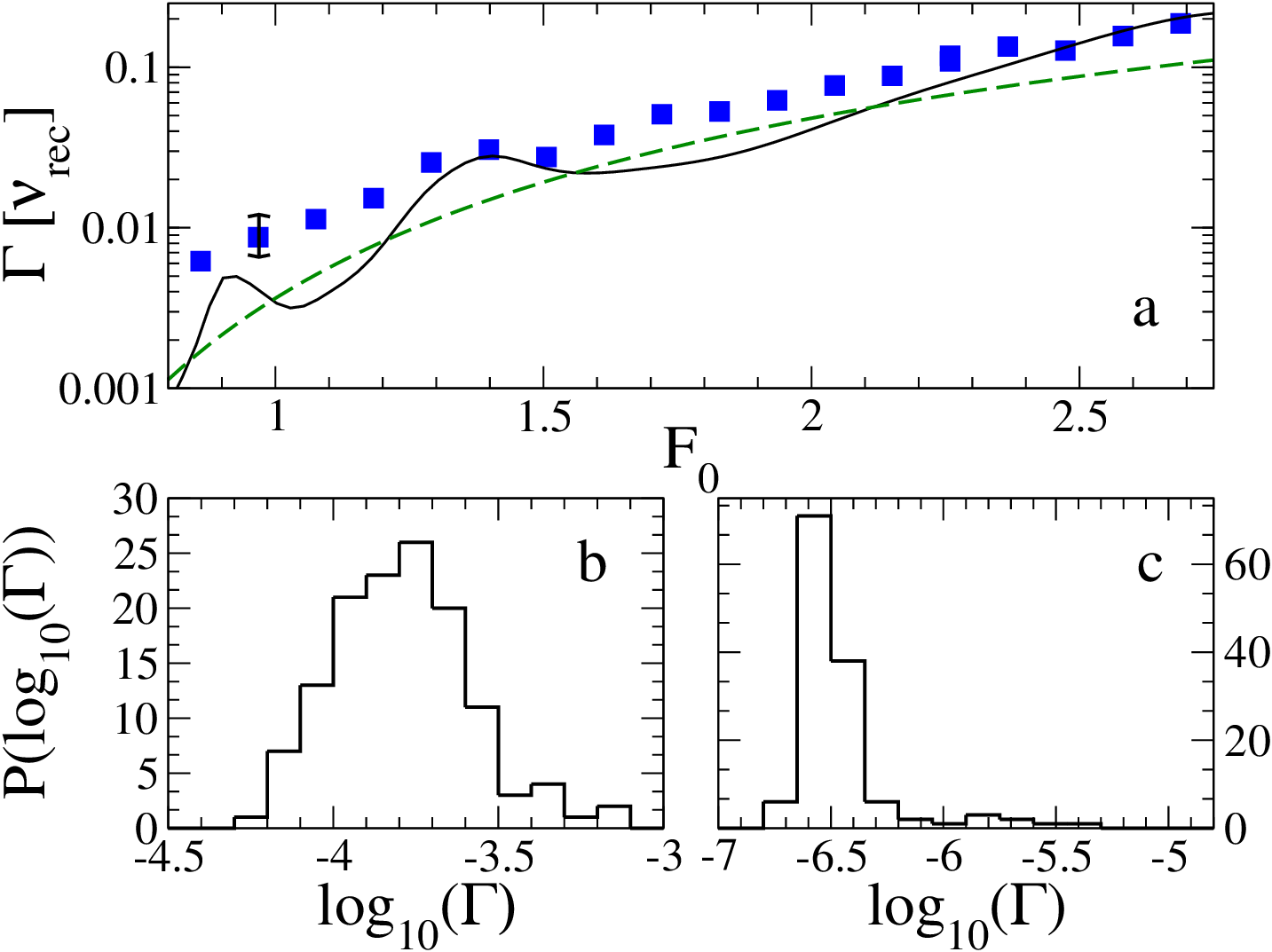}
\caption{\label{fig:3} 
($a$) same theoretical data as in the upper graph of figure~\ref{fig:2} with measured tunnelling rates
for a small nonlinearity $C \approx 0.025$ (\fullsquare). 
The mean-field, repulsive nonlinearity washes
out the RET peak structures of figure~\ref{fig:2}, and this can be used as an additional experimental
handle to globally enhance the rates. ($b$) and ($c$) show numerical data from a Bose-Hubbard
model of 7 atoms in 6 potential wells for $V_0=3$ (fixing the hopping constant $J\approx 0.22$), interaction
constant $U=0.2$, and $F_0=0.47$ ($b$) or $F_0=0.16$ ($c$). 
The broad distribution at the larger force arises from the regular
Bloch oscillation dynamics of the atoms. The distribution in ($c$) shows that there is a small number
of preferred channels to tunnel to the first excited levels originating from the quantum chaotic
motion of the atoms along the wells (leading e.g. to interaction-induced decoherence of the
Bloch oscillations \cite{BK2003}).
}
\end{figure}

Our experimental setup is well-modelled by the following three-dimensional Gross-Pitaevskii equation
\cite{M2001,Wim2005}:
\begin{eqnarray}
&{\rm i} \hbar \frac{\partial }{\partial t}\psi (\vec{r},t) =
\left[-\frac{\hbar^2}{2M}\nabla ^2 + \frac{1}{2}M  \left( \omega_x x^2 + \omega_y y^2 +
\omega_z z^2 \right) + V \sin^2\left(\frac{\pi x}{d_L}\right) + F x 
\right. \nonumber \\ & \left.
+ g N \left| \psi(\vec{r},t) \right|^2
\right] \psi(\vec{r},t) \;.
\label{eq:GP}
\end{eqnarray}
$\psi(\vec{r},t)$ represents the wave function of Bose-Einstein condensate, and the various potential
terms can be used to engineer the dynamics and the tunnelling decay of the quantum gas. The three
frequencies $\vec{\omega}$ describe the longitudinal and the transversal trap frequencies which can be
varied to change the density of the condensate within the wells. In such a way, we can adapt the
effective nonlinearity $ g N \left| \psi(\vec{r},t) \right|^2$, where $g=4\pi \hbar^2 a_s/M$, with
the $s$-wave scattering length $a_s$ and the total number of atoms $N$.
Not only can we control the many-body evolution of the condensate by adapting the atomic density, but
we are also able to guide the dynamical behaviour by varying the potential depth $V$ and the force $F$.
As a consequence, we have all possible handles to switch between ``fast'' and ``slow'' tunnelling escape
of the condensate using in particular the RET effect sketched in figure~\ref{fig:1}.
The optimal extraction of the tunnelling rates from the experimentally observable momentum distributions
of the condensate at different evolution times is described in full detail in a recent theoretical 
paper~\cite{Wim2005}. Here we restrict to show the main experimental results which can be found in the 
figures~\ref{fig:2} and \ref{fig:3}.

Figure~\ref{fig:2} shows the decay rate for a condensate of Rubidium atoms initially populating just the 
energetically lowest state within each well of figure~\ref{fig:1}, for small atomic density, i.e. for 
practically vanishing nonlinearity. The solid line represents the theoretical prediction for $g=0$ in
eq.~(\ref{eq:GP}). The dashed line shows the expected decay derived from the perturbative Landau-Zener 
formula $ \Gamma_{\rm LZ}=(\nu_{\rm rec} F_0) \exp(-\pi^2V_0^2/32F_0) $, with $V_0\equiv V/E_{\rm rec}$
and $F_0 \equiv F d_L/E_{\rm rec}$, $E_{\rm rec}\equiv \hbar^2 \pi^2/(2Md_{\rm L}^2)$ being the recoil
energy and $d_{\rm L}$ the lattice constant \cite{holt}.
The modulation originating from RET on top of the simple Landau-Zener prediction can, in principle,
reach several orders of magnitude \cite{GKK}, and an example for the enhancement by a 
factor of $1.5$ is shown in the lower panel of figure~\ref{fig:2} (directly 
showing the ratio between the measured rate and $ \Gamma_{\rm LZ} $).

Figure~\ref{fig:3} (a) highlights the impact of the mean-field nonlinearity on the tunnelling rates. The
effective nonlinearity parameter is conveniently expressed by $C \equiv gn_0/(8E_{\rm rec})$, with
the peak density of the condensate $n_0$, and $C$ can be directly estimated from independent
measurements \cite{M2001}. In accordance with the theoretical prediction of \cite{Wim2005}, the
repulsive interatomic interactions have two main effects. First, they tend to enhance the decay rate
for all values of $F$. Second, they wash out the RET peak structure due to the
different scaling of the tunnelling rates with the nonlinear parameter $C$ at the peak maximum
and in the region where the Landau-Zener formula applies \cite{RET_12}. In this latter regime, the scaling is
found to be linear in $C$ \cite{RET_12}, as expected from perturbation theory, which, on the other hand, 
does not apply at the peak maxima where energy levels are degenerate.

While the experiment can access the decay rates for short times (up to about ten Bloch periods), for
longer times, the density dependent decay naturally leads to a nonexponential scaling of the survival
probability of the condensate in the open Wannier-Stark system. Such a nonexponential tunnelling might
be observable in future experiments which could either realize larger nonlinearities (we estimate from
numerical data $C > 0.25$ \cite{Wim2005}), or be sensitive enough to measure the long-time behaviour of 
the survival probability close to a RET peak \cite{SW}. 

We would like to emphasise that the presence of atom-atom interactions in ultracold atomic gases does not
hinder possible applications of the here observed RET effect. On the contrary, the effective interaction
can be controlled and used as an additional handle to enhance (for repulsive interactions) or to
decrease (for attractive interactions) the quantum tunnelling in an extended potential \cite{Wim2005}.
In the here presented experiments, the interactions can be controlled by the nonlinearity parameter
$C \approx 0\ldots 0.1$ \cite{M2001,RET_12}. 

Other experimental setups can reduce the density of atoms
to the order of one atom per lattice well \cite{bloch}, which makes necessary a true many-body description
of the problem going beyond the mean-field equation (\ref{eq:GP}). An adequate description of the
many-body dynamics is complicated but possible for small lattices and filling factors of the order one
\cite{BK2003}. Based on the Bose-Hubbard model, we could deduce that the tunnelling between the 
lowest and the first excited Wannier-Stark levels depends indeed crucially on the dynamical properties of 
the many-body problem. The latter are controlled again by the Stark force $F$. If the many-body
dynamics shows signatures of quantum chaos, we have found that the system has preferred decay channels,
whilst for regular (i.e. close to integrable motion) of the atoms along the lattice, the distribution
of the tunnelling rates is relatively broad \cite{T_diplom}. The latter numerical results were obtained
from a single-band Bose-Hubbard model for which the decay to the excited energy levels on each lattice site
was perturbatively included \cite{T_diplom}. The corresponding distributions are collected
in the panels ($b$) and ($c$) of figure~\ref{fig:3} as examples of regular and chaotic motion, respectively.

\section{Conclusions} 
\label{concl}\medskip

In summary, our combined theoretical and experimental analysis of 
the tunnelling decay of ultracold atoms in periodic potentials show that quantum tunnelling
can be easily controlled by adapting the parameters of the system. Experiments with Bose-Einstein
condensates in optically produced potentials allow us the required degree of control making them
an ideal playground to study solid-state models such as the Wannier-Stark system.

In consequence, we can propose the RET effect as an experimental handle to engineer quantum
tunnelling, and as a possible starting point for future applications. We may think of
diode-like switches between spatially separated condensates, or continue the route to study true
many-body tunnelling for which dynamical properties of the system as a whole (including strong
interatomic interactions \cite{Pon2006}) and possible external couplings (e.g. to leads or to heat bathes 
\cite{livi}) conspire to lead to new interesting physical phenomena.

\ack
S W thanks with pleasure the organiser Thomas Elze for his kind invitation to  DICE2006.
We are grateful to Andreas Buchleitner, Matteo Cristiani, Hans Lignier, Andrey Kolovsky,
Peter Schlagheck, Carlo Sias, Andrea Tomadin, Dirk Witthaut and Alessandro Zenesini for assistance and 
many useful discussions. Our work is supported by the EU (OLAQUI), MIUR (PRIN), CNISM,
and the Alexander von Humboldt Foundation (Feodor-Lynen Program).


\section*{References}

\medskip
\begin{thebibliography}{99}

\bibitem{qm}
Razavy M 2003 {\it Quantum Theory of Tunneling} (World Scientific, Singapore)

\bibitem{heller}
Davis M J and Heller E J 1981 {\it J. Chem. Phys.} {\bf 75} 246;
Creagh S 1998 in {\it Tunneling in Complex Systems} edited by Tomsovic S (World Scientific, Singapore) p. 1

\bibitem{dyntun}
N\"ockel J U and Stone A D, 1997 {\it Nature} {\bf 385} 45; 
Zakrzewski J, Delande D and Buchleitner A 1998 {\it Phys. Rev. E} {\bf 57} 1458;
Steck D A, Oskay W H and Raizen M G 2001 {\it Science} {\bf 293} 274;
Hensinger W K {\it et al} 2001 {\it Nature} {\bf 412} 52;
Sheinman M, Fishman S, Guarneri I and Rebuzzini L 2006 {\it Phys. Rev. A} {\bf 73} 052110;
Wimberger S, Schlagheck P, Eltschka C and Buchleitner A 2006 {\it Phys. Rev. Lett.} {\bf 97} 043001

\bibitem{binnig}
Binnig G and Rohrer H 1987 {\it Rev. Mod. Phys.} {\bf 59} 615; 1999 {\it Rev. Mod. Phys.} {\bf 71} S324

\bibitem{koelle}
Koelle D, Kleiner R, Ludwig F, Dantsker E and Clarke J 1999 {\it Rev. Mod. Phys.} {\bf 71} 631

\bibitem{esaki1974}
Esaki L  1974 {\it Rev. Mod. Phys.} {\bf 46} 237

\bibitem{esaki1986}
Esaki L 1986 \textit{IEEE Journal Quant. Electr.} {\bf QE-22} 1611

\bibitem{chang}
Chang L L, Esaki L and Tsu R 1974 \textit{Appl. Phys. Lett.} \textbf{24} 593 

\bibitem{chang_book}
Chang L L, Mendez E E and Tejedor C (eds) 1991 \textit{Resonant
Tunneling in Semiconductors} (Plenum, New York); 
Tarucha S and Ploog K 1988 {\it Phys. Rev. B} {\bf 38} 4198; 
Lievescu G, Fox A M, Miller D A, Sizer T, Know W H, Gossard A C and Englisch J H 1989  
{\it Phys. Rev. Lett.}  {\bf 63}  438; Spielman I B, Eisenstein J P, Pfeiffer L N and West K W
2000  {\it Phys. Rev. Lett.}  {\bf 84} 5808; Rosam B, Leo K, Gl\"uck M, Keck F, Korsch H J, Zimmer F and
K\"ohler K 2003 {\it Phys. Rev. B}  {\bf 68} 125301.

\bibitem{peik}
Dahan M B, Peik E, Reichel J, Castin Y and Salomon C 1996
\textit{Phys.\ Rev.\ Lett.} \textbf{76} 4508; 
Wilkinson S R, Bharucha C F, Madison K W, Niu Q and Raizen M G 1996 {\it Phys. Rev. Lett.} {\bf 76} 4512 

\bibitem{GKK}
Gl\"uck M, Kolovsky A R and Korsch H J 2002 \textit{Phys. Rep.} \textbf{366} 103

\bibitem{WSM}
Wimberger S, Schlagheck P and Mannella R 2006 {\it J. Phys. B: At. Mol. Opt. Phys.} {\bf 39} 729

\bibitem{avron} Avron J E 1982 \textit{Ann. Phys.} \textbf{143} 33

\bibitem{WGWK}
Witthaut D, Graefe E M, Wimberger S and Korsch H J 2006  {\it Phys. Rev. A} at press

\bibitem{bha} 
Bharucha C F, Madison K W, Morrow P R, Wilkinson S R, Sundaram B and Raizen M G 
1997 \textit{Phys. Rev. A} \textbf{55} R857

\bibitem{M2001}
Morsch O, M\"uller J H, Cristiani M, Ciampini D and Arimondo E 2001
\textit{Phys. Rev. Lett.} \textbf{87} 140402


\bibitem{Wim2005}
Wimberger S, Mannella R, Morsch O, Arimondo E, Kolovsky A R and Buchleitner A 2005 \textit{Phys.
Rev. A} \textbf{72} 063610.

\bibitem{holt}
Holthaus M 2000 {\it J. Opt. B: Quant. S. Opt.} {\bf 2} 589

\bibitem{RET_12}
Sias C, Zenesini A, Lignier H, Wimberger S, Ciampini D, Morsch O and Arimondo E
2006 in preparation

\bibitem{SW}
Schlagheck P and Wimberger S 2006 {\it Appl. Phys. B} at press

\bibitem{bloch}
Greiner M, Mandel O, Esslinger T, H\"ansch T W and Bloch I 2002 {\it Nature} {\bf 415} 39;
Mandel O, Greiner M, Widera A, Rom T, H\"ansch T W and Bloch I 2003 {\it Nature} {\bf 425} 937;
St\"oferle T, Moritz H, Schori C, K\"ohl M and Esslinger T 2004 {\it Phys. Rev. Lett.} {\bf 92} 130403; 
F\"olling S,  Widera A, M\"uller T, Gerbier F and Bloch I 2006 {\it Phys. Rev. Lett.} {\bf 97} 060403

\bibitem{BK2003}
Buchleitner A and Kolovsky A R 2003 {\it Phys. Rev. Lett.} {\bf 91} 253002

\bibitem{T_diplom}
Tomadin A 2006 {\it Quantum chaos with ultracold atoms in optical lattices} 
(Universit\`a degli Studi di Pisa, Master's Thesis)

\bibitem{Pon2006}
Ponomarev A V, Madro\~nero J, Kolovsky A R and Buchleitner A 2006 
{\it Phys. Rev. Lett.} {\bf 96} 050404

\bibitem{livi}
Livi R, Franzosi R and Oppo G L 2006 {\it Phys. Rev. Lett.} {\bf 97} 060401;
see also the contributions of Mahler G and Lombardo F C {\em et al} 
in this issue of the DICE2006 proceedings

\end{thebibliography}
\end{document}